# Multifractal Measures on Small-World Networks

Kyungsik Kim,* K. H. Chang,[a] S. M. Yoon,[b] C. Christopher Lee,[c] J. S. Choi

*Department of Physics, Pukyong National University, Pusan 608-737, Korea*

[a]*Forecast Research Laboratory, Meteorological Research Institute, KMA, Seoul 156-720, Korea*

[b]*Division of Economics, Pukyong National University, Pusan 608-737, Korea*

[c]*Department of Business Administration, College of Business, Central Washington University, WA 98926, USA*

## ABSTRACT

We investigate the multifractals of the normalized first passage time on one-dimensional small-world network with both reflecting and absorbing barriers. The multifractals is estimated from the distribution of the normalized first passage time characterized by the random walk on the small-world network with three fractions of edges rewired randomly. Particularly, our estimate is the fractal dimension $D_0$ = 0.917, 0.926, 0.930 for lattice points $L = 80$ and a randomly rewired fraction $p = 0.2$. The numerical result is found to disappear multifractal properties in the regime $p > p_c = 0.3$, where $p_c$ is the critical rewired fraction.

For last few years, many interests have been concentrated on small-world and scale-free network models [1] in the physical, biological, social and technological networks. The small-world network, proposed by Watts and Strogatz [2], has played a crucial role in connected properties represented by single component graphs. The real-world models are indeed characterized by the small world and clustering property, e.g., social nets [3], the internet [4], document retrieval in www[5], scientific cooperation relations [6], social networks of acquaintances [7] and of collaborations [8]. The static and dynamic behaviors extensively have studied on small-world networks, and the prominent topics for small-world networks really have a direct application in statistical mechanics and polymer physics [9].

On the other hand, one of stochastic processes for regular and disordered systems includes the random walk theory [10], and a line of development of importance for the advancement of the random walk is the continuous-time random walk theory, formerly introduced by Montroll and Weiss [11], which substantially elaborated on the transition probability and the distribution of the pausing times [12]. To describe quantitatively hopping of the continuous-time random walk problem, one has to know how to calculate both the transition probability and random pausing time of the random walker between lattice points. In particular, the transition rate in charge conductions of the disordered solid has depended on the spatial separation between the localized centers and on the energy difference between the initial and final state [13]. Until now, the continuous-time random walk theory mainly has studied in natural and social sciences, and among the outstanding topics, reaction and strange kinetics [14], fractional diffusion equations [15], random networks, earthquake modeling, hydrology, and financial options [16] have been the focus of attention to many researchers.

The mean first passage time for the random walk theory is the statistical quantity that defines the average time arriving at the absorbing barrier for the first time. Particularly, Sinai model [17,18] has discussed the first passage time related to the random barrier with the absorbing and reflecting barriers. The Sinai model and other previous works [19] were studied for the mean and mean square displacements dependent anomalously on time. Many stochastic processes on one-dimensional lattice have argued with the mean first passage time, for the specific cases of the transition probability such as chaotic orbits of the logistic and Kim-Kong maps [20]. Recently, Kim and Kong [21] have used the box-counting method to estimate the generalized dimension and the scaling exponent for the mountain height and the sea-bottom depth.

In this paper, we consider a randomly rewired fraction $p$ of one-dimensional small-world networks, so that choosing $p \in [0, 1]$ allows to interpolate between the regular ($p = 0$) and random ($p = 1$) networks. The purpose of this paper is to apply the small-world graph theory to the random walk on one-dimensional lattice with both reflecting and absorbing barriers. The multifractals from the normalized first passage time is estimated on small-world networks with three fractions of edges rewired randomly, i.e. $p = 0.1, 0.2, 0.3$ and compare our result with that on the regular lattice.

In this study, our model considers the random walk on one-dimensional lattice with a randomly rewired fraction of edges. Then, the transition probability $p(l)$ is introduced:

$$p(l) = \alpha_i \, \delta_{l,i+a} + \beta_i \, \delta_{l,i-b} , \qquad (1)$$

where $\alpha_i = p_i / k_i$ ($\beta_i = q_i / m_i$) is the transition probability that the random walker jumps to a right (left) nearest-neighboring lattice point $i + a$ ($i - b$) from a lattice point $l = i$, and $a$ and $b$ are positive integers. The integer $k_i$ ($m_i$) is the number of a right-sided (left-sided) nearest-neighboring lattice points of a site $l = i$. The normalization condition $\sum_l p(l) = 1$ is conserved, and the transition probability conservation is also expressed as $p_i + q_i = 1$. At a discrete time $t$, the probability density $P_i(t)$ is given by

$$P_i(t) = \beta_{i+a} \, P_{i+a}(t-1) + \alpha_{i-b} \, P_{i-b}(t-1) , \qquad (2)$$

where $P_i(t)$ is the probability density for the random walker to exist at a lattice point $l = i$ at time $t$, and $\alpha_{i-b} = p_{i-b}/k_{i-b}$ ($\beta_{i+a} = q_{i+a}/m_{i+a}$) is the transition probability that the random walker jumps to a right (left) nearest-neighboring lattice point from a lattice point $l = i - b$ ($l = i + a$). We also introduce the normalized first passage time to arrive at the absorbing barrier after starting from an arbitrary site. Using the generating function technique [15], the mean first passage time $<t>$ in a randomly rewired one-dimensional lattice is represented in terms of

$$\langle t \rangle = \sum_{k=1}^{N-1} \frac{1}{\alpha_k} + \sum_{k=0}^{N-2} \frac{1}{\alpha_k} \sum_{i=k+1}^{N-1} \prod_{j=k+1}^{i} \beta_j . \qquad (3)$$

Here, one particle moves randomly to a right (left) nearest-neighboring lattice point with $\alpha_j$ ($\beta_j$) after one step after starting from lattice pont $l = j$. From Eq. (1), the normalized first passage time is defined by

$$t_{nt} = \frac{t_n - t_{min}}{t_{max} - t_{min}}, \qquad (4)$$

where $t_n$ is the first passage time arriving at the absorbing barrier after time $t$, and $t_{max}$

($t_{min}$) is the maximum (minimum) value of the first passage time arriving at the absorbing barrier.

To analyze the multifractal feature [22] for the motion of a random walker on small-world network, we review the generalized dimension and the spectrum of our interest in the multifractal formalism. We suppose that the normalized first passage time is divided into $M(\varepsilon)$ cells of size, $\varepsilon = 1/M(\varepsilon)$. For a given normalized first passage time $t_{nt}$ of length $N$, let $p_i = n_i/N$ denote the fractional number of points in the $i$ - th cell, $[(i-1)\varepsilon, i\varepsilon]$. Then, the generalized dimension in multifractal structures is represented as

$$D_q = \lim_{\varepsilon \to 0} \frac{\ln \sum_i n_i p_i^q}{(q-1)\ln \varepsilon}. \qquad (5)$$

The spectra $f_q$ and $\alpha_q$ are related to $D_q$ via the following Legendre transform:

$$f_q = \alpha_q - (q-1)D_q \qquad (6)$$

and

$$\alpha_q = \frac{d}{dq}[(q-1)D_q]. \qquad (7)$$

In this scheme, we will make use of Eqs. (5)-(7) to compute the multifractal measures for the small-world network, and these mathematical techniques lead us to more general results. The fractal dimension obtained from the multifractal measures can be compared numerically with other regular cases.

To estimate numerically the multifractals on one-dimensional small-world network, we mainly concentrate on the generalized dimension and the spectrum for the normalized first passage time. For all lattice points $L$, the reflecting barrier is assumed to locate on lattice point $l = 1$, while the absorbing barrier is at $l = L$. We can analyze Eq. (4) from the first passage time arriving at the absorbing barrier for the first time, after a random walker starts from an initial point $l = 0.1L$. The computer simulations are carried out according to Eq. (4) for three lattice sizes $L = 60, 80, 100$. We average over at least $10^3$ independent configurations and have $10^4$ random walkers per configuration.

To carry out the random walk on one-dimensional small-world network, we only restrict ourselves to three rewired fractions of edges $p = 0.1, 0.2, 0.3$. We assume that the random walker jumps to a right (left) nearest-neighboring lattice point from a lattice point $l = i$ with the transition probability $p_i/k_i$ ($q_i/m_i$), where $k_i$ ($m_i$) is the number of a right (left) nearest-neighboring lattice points of a site $l = i$. As we perform computer simulations until the random walker arrives at the absorbing barrier for the first time, we can obtain the multifractals of the normalized first passage time on the one-dimensional small-world network. In Figs. 1 and 2, we show the plot of the generalized dimension and spectra $f_q$ and $a_q$, and our estimate is the fractal dimension $D_0$=0.917, 0.926, 0.930 for lattice points $L = 80$ and a randomly rewired fraction $p = 0.2$. When the randomly rewired fraction is larger than 0.3, the calculated result of our small-world network is found to disappear mulfractal properties in practice. Here, $p_c = 0.3$ is the critical rewired fraction which has the so-called marginal value of $p$. The other numerical result for the fractal dimension, the generalized dimension, and spectra on our small-world network is summarized in Table 1.

Furthermore, we report the numerical results for multifractals on a regular one-dimensional lattice. Fig. 3 shows the plot of spectra $f_q$ and $a_q$ for various system sizes. The generalized dimension for system sizes $L = 60, 80, 100$ is depicted in Fig. 4, and our estimate is the fractal dimension $D_0 = 0.996, 0.998, 0.999$. The other numerical result for the generalized dimension and spectrua is shown in Table 2.

In summary, we have studied the random walk dynamics on one-dimensional small-world network with both reflecting and absorbing barriers. We have found numerically the generalized dimension and the spectrum of our interest from the multifractal formalism, in order to analyze the multifractal feature for the random walker on small-world network. As summarized in Table 1, the value of the fractal dimension for three cases of randomly rewired fraction is found to be significantly different from that near one on a regularly one-dimensional lattice. In the future, we will extensively investigate the multifractals dealt with complicated stochastic models on small-world network and compare in detail with calculations performed in other models.

### ACKNOWLEDGMENT

This work was supported by Korea Research Foundation Grant(KRF-2003-002-B00032).

# Table Captions

| Lattice points $L$ | Rewired fraction $p$ | $D_0$ | $D_{-\infty}$ | $D_{+\infty}$ | $a_{-\infty}$ | $a_{+\infty}$ | $f_{-\infty}$ | $f_{+\infty}$ |
|---|---|---|---|---|---|---|---|---|
| 60 | 0.1 | 0.922 | 1.443 | 0.696 | 1.456 | 0.687 | 0.586 | 0.084 |
|    | 0.2 | 0.930 | 1.580 | 0.673 | 1.596 | 0.665 | 0.539 | 0.096 |
|    | 0.3 | 0.920 | 1.677 | 0.707 | 1.694 | 0.699 | 0.507 | 0.142 |
| 80 | 0.1 | 0.917 | 1.406 | 0.696 | 1.418 | 0.687 | 0.596 | 0.077 |
|    | 0.2 | 0.926 | 1.539 | 0.676 | 1.553 | 0.667 | 0.553 | 0.093 |
|    | 0.3 | 0.930 | 1.635 | 0.672 | 1.652 | 0.664 | 0.525 | 0.139 |
| 100 | 0.1 | 0.917 | 1.383 | 0.698 | 1.394 | 0.690 | 0.606 | 0.076 |
|    | 0.2 | 0.925 | 1.509 | 0.678 | 1.523 | 0.670 | 0.566 | 0.090 |
|    | 0.3 | 0.930 | 1.600 | 0.664 | 1.615 | 0.656 | 0.535 | 0.099 |

**Table 1**  Summary of values of the fractal dimension, the generalized dimension, and the scaling exponents on randomly rewired one-dimensional lattice consisting of three values of lattice points $L$ = 60, 80, 100.

| Lattice points $L$ | $D_0$ | $D_{-\infty}$ | $D_{+\infty}$ | $a_{-\infty}$ | $a_{+\infty}$ | $f_{-\infty}$ | $f_{+\infty}$ |
|---|---|---|---|---|---|---|---|
| 60 | 0.996 | 1.142 | 0.894 | 1.168 | 0.889 | 0.310 | 0.636 |
| 80 | 0.998 | 1.172 | 0.885 | 1.195 | 0.880 | 0.412 | 0.648 |
| 100 | 0.999 | 1.163 | 0.921 | 1.188 | 0.916 | 0.354 | 0.663 |

**Table 2**  Summary of values of the fractal dimension, the generalized dimension, and the scaling exponents for three values of lattice points $L$ = 60, 80, 100 on one-dimensional lattice with reflecting and absorbing barriers.

# Figure Captions

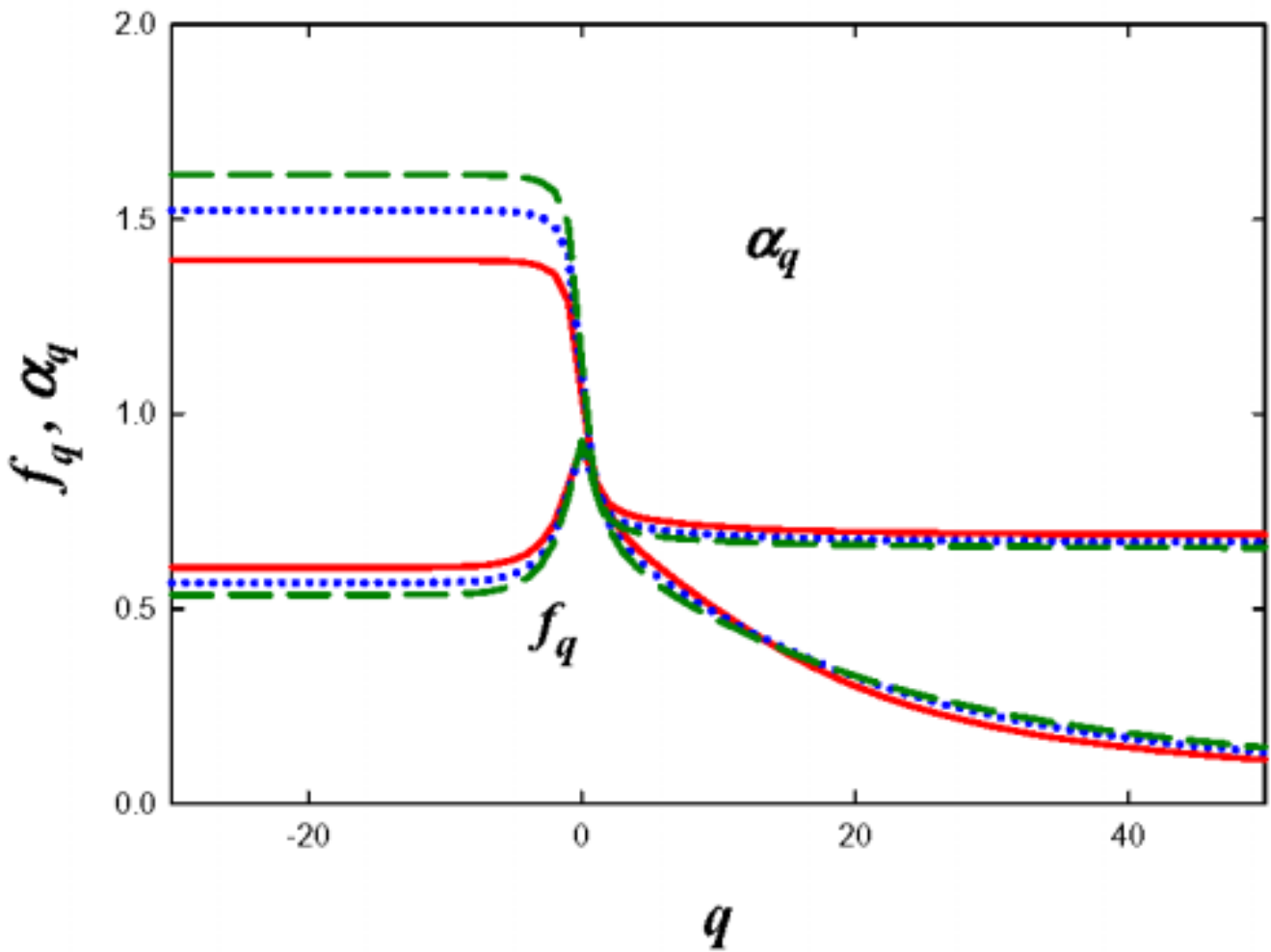

**Fig. 1** Plot of spectra $f_q$ and $a_q$ for three values of lattice points $L = 100$ and the rewired fraction of edges $p = 0.1$(solid line), 0.2(dot line), 0.3(dashed line) on one-dimensional small-world networks with reflecting and absorbing barriers.

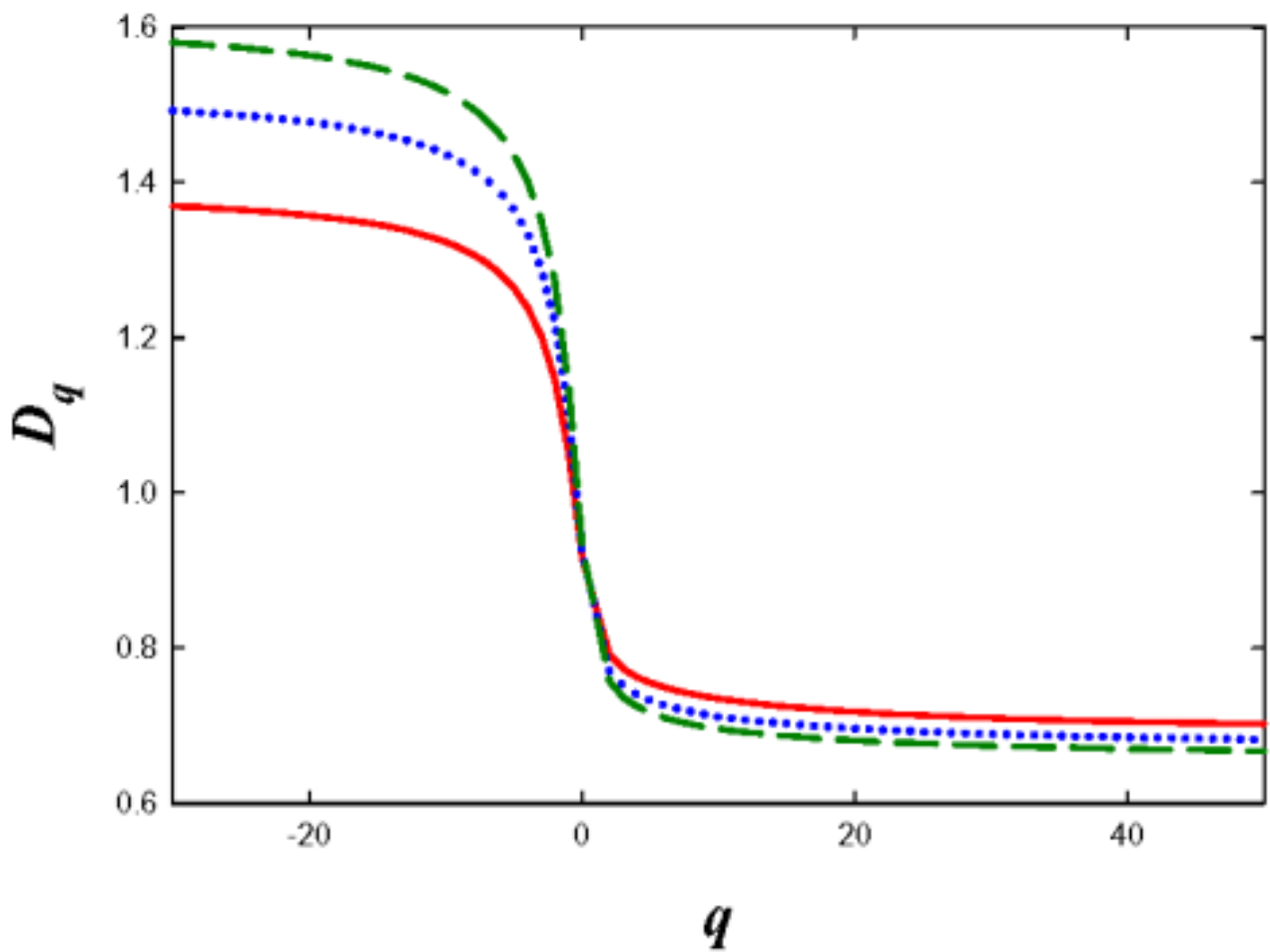

**Fig. 2** Plot of the generalized dimension $D_q$ versus $q$ for three values of lattice points $L$ = 100 and the rewired fraction of edges $p$ = 0.1(solid line), 0.2(dot line), 0.3(dashed) on one-dimensional small-world networks.

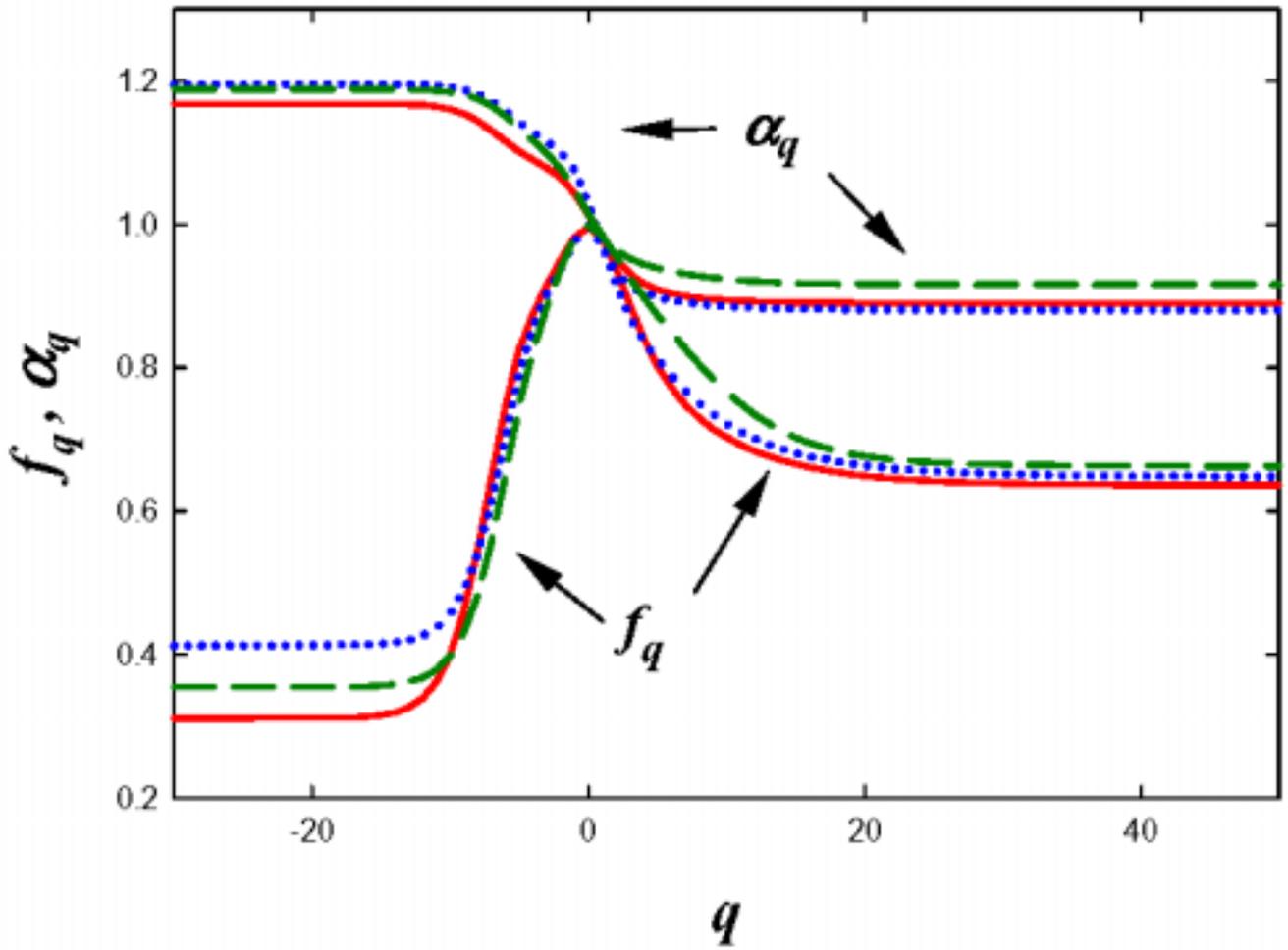

**Fig. 3** Plot of spectra $f_q$ and $a_q$ for three values of lattice points $L = 60$(solid line), 80 (dot line), 100(dashed line) on one-dimensional lattice with reflecting and absorbing barriers.

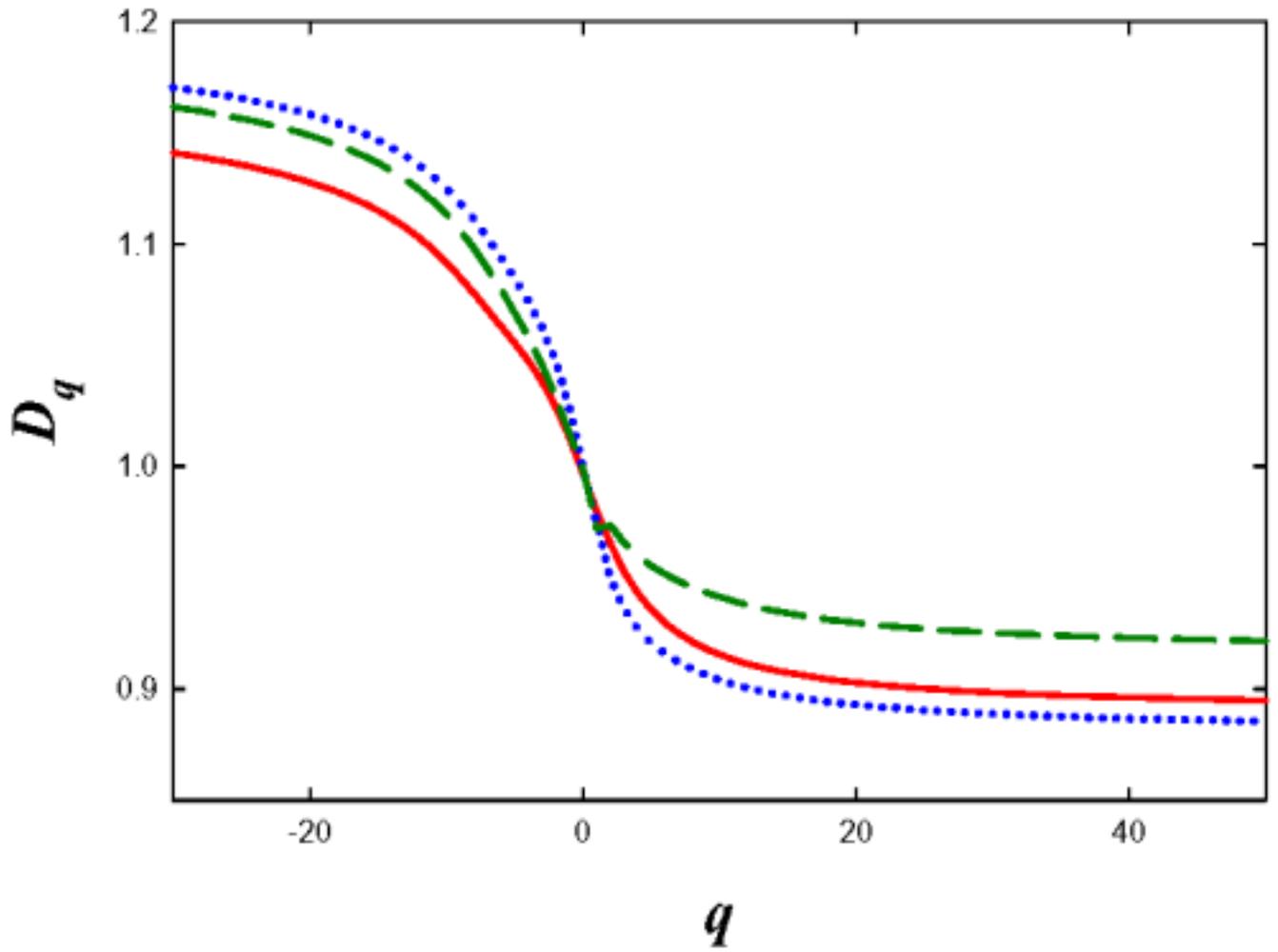

**Fig. 4** Plot of the generalized dimension $D_q$ versus $q$ for three values of lattice points $L$=60(solid line), 80(dot line), 100(dashed line) on one-dimensional lattice with reflecting and absorbing barrier.